# Indicative Surfaces for Crystal Optical Effects

R.Vlokh, O.Mys and O.Vlokh

Institute of Physical Optics, 23 Dragomanov St., 79005 Lviv,
Ukraine, E-mail: vlokh@ifo.lviv.ua

**Abstract:** This paper has mainly a pedagogical meaning. Our aim is to demonstrate a correct general approach for constructing indicative surfaces of higher-rank tensors. We reconstruct the surfaces of piezo-optic tensor for β-BaB$_2$O$_4$ and LiNbO$_3$ crystals, which have been incorrectly presented in our recent papers.

**Introduction**

Many works have been devoted to analysis of anisotropy of piezooptic (PO) effect in crystals on the basis of construction of so-called indicative surfaces (see [1-5] and our previous studies [6,7]). Unfortunately, the indicative surfaces in those studies have been constructed, using the inconsistent formula

$$\pi'_{ijkl} = \alpha_{im}\alpha_{jn}\alpha_{kp}\alpha_{lt}\pi_{mnpt}, \qquad (1)$$

where $\pi'_{ijkl}$ means the PO tensor written in the current (or 'new') Cartesian coordinate system, $\pi_{mnpt}$ the PO tensor written in the 'old' Cartesian coordinate system and $\alpha_{im}, \alpha_{jn}, \alpha_{kp}, \alpha_{lt}$ the matrix components of directional cosines between the 'old' and 'new' systems. It is necessary to note that Eq. (1) is often used for rewriting tensors in different Cartesian systems or, in some special cases, for constructing the so-called representation surfaces (see, e.g., [8]), though not the indicative ones. In this paper, we present the relation for the indicative surfaces of higher-order crystal optical effects, specifically for the PO one, and reconstruct in this way the surfaces presented in our previous works.

**Results and Discussion**

Using a spheric coordinate system $(R, \Theta, \varphi)$, one can represent the relation for the indicative surfaces of symmetric tensors of higher ranks as follows [9]:

$$R(\Theta, \varphi) = T_{i_1...i_p} n_{i_1}...n_{i_p}, \qquad (2)$$

where $R$ is the module of the spheric coordinate system, $T_{i_1...i_p}$ the tensor of a rank $p$ and $n_{i_1}...n_{i_p}$ the transformation relation between Cartesian and spheric coordinates ($n_1 = \sin\Theta\cos\varphi, n_2 = \sin\Theta\sin\varphi, n_3 = \cos\Theta$). Applying Eq. (2) to the PO effect, we find easily the relation for the surface of the PO tensor:

$$R(\Theta, \varphi) = \pi_{ijkl} n_i n_j n_k n_l. \qquad (3)$$

Eq. (2) has been successfully used while constructing the indicative surfaces of second-rank axial gyration tensor, elastic module tensor, etc. (see, e.g., [9,10]). Since for the case of PO effect $R \sim \pi_{ijkl} \sim \frac{\Delta n}{\sigma_{kl}}$, the $R$ value would reflect in some cases anisotropy of the induced increment of refractive indices $\Delta n$ due to the action of mechanical stresses $\sigma_{kl}$ in different directions. Moreover, it is possible to construct a part of the indicative surface if one takes an interest in this. For example, in the case of PO anisotropy induced by a chosen component of mechanical stress tensor, it is reasonable to take into account only one column that corresponds to a given stress component.

Let us construct the indicative surface of the PO tensor for LiNbO$_3$ and $\beta$-BaB$_2$O$_4$ crystals. This would be instructive enough for the illustrative purposes, as well as for making necessary corrections to our previous construction procedure [6,7]. Both LiNbO$_3$ and $\beta$-BaB$_2$O$_4$ crystals belong to the same point symmetry group *3m*. Then the equation of the indicative surface takes the same form for those crystals:

$$R(\theta,\varphi) = \pi_{11}\sin^4\theta + (\pi_{13} + \pi_{31} + 2\pi_{44})\sin^2\theta\cos^2\theta + \pi_{33}\cos^4\theta + $$
$$+(2\pi_{41} + \pi_{14})\sin^3\theta\cos\theta\sin3\varphi \qquad (4)$$

Using the known values of PO coefficients for β-BaB$_2$O$_4$ crystals, $\pi_{11} = (-1.7 \pm 0.15)\,pm^2/N$, $\pi_{12} = (-1.35 \pm 0.07)\,pm^2/N$, $\pi_{13} = (1.75 \pm 0.23)\,pm^2/N$, $\pi_{31} = (-1.6 \pm 0.15)\,pm^2/N$, $\pi_{33} = (3.7 \pm 0.37)\,pm^2/N$, $\pi_{14} = (-2.0 \pm 0.8)\,pm^2/N$, $\pi_{41} = (-2.03 \pm 0.07)\,pm^2/N$, $\pi_{44} = (-26.3 \pm 0.9)\,pm^2/N$, one can represent Eq. (4) as

$$R(\theta,\varphi) = (-1.7\sin^4\theta - 52.45\sin^2\theta\cos^2\theta + 3.7\cos^4\theta - 6.06\sin^3\theta\cos\theta\sin3\varphi)\times 10^{-12}. \qquad (5)$$

In case of LiNbO$_3$ crystals we have $\pi_{11} = -47.7\,pm^2/N$, $\pi_{12} = 0.11\,pm^2/N$, $\pi_{13} = 2\,pm^2/N$,

$\pi_{31} = 0.47\,pm^2/N$, $\pi_{33} = 1.6\,pm^2/N$, $\pi_{14} = 0.7\,pm^2/N$, $\pi_{41} = -1.9\,pm^2/N$, $\pi_{44} = 0.21\,pm^2/N$ and so we obtain

$$R(\theta,\varphi) = (-0.47\sin^4\theta + 2.89\sin^2\theta\cos^2\theta + 1.6\cos^4\theta - 3.1\sin^3\theta\cos\theta\sin3\varphi)\times 10^{-12}. \qquad (6)$$

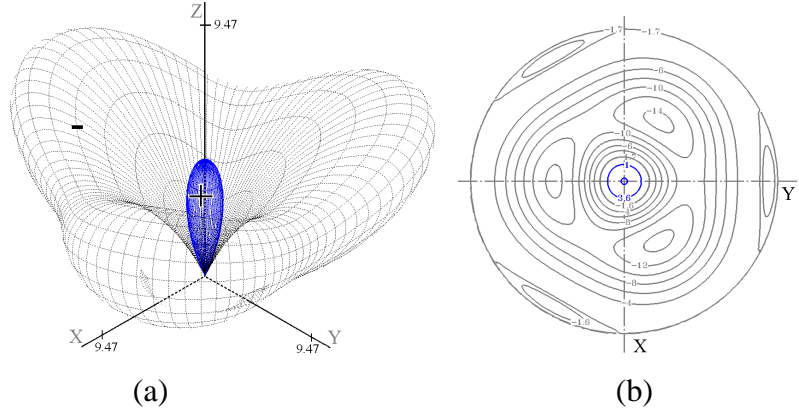

(a)        (b)

Fig. 1. Half the indicative surface of the PO tensor (a) and its stereographic projection (b) for β-BaB$_2$O$_4$ crystals (in the units of $pm^2/N$).

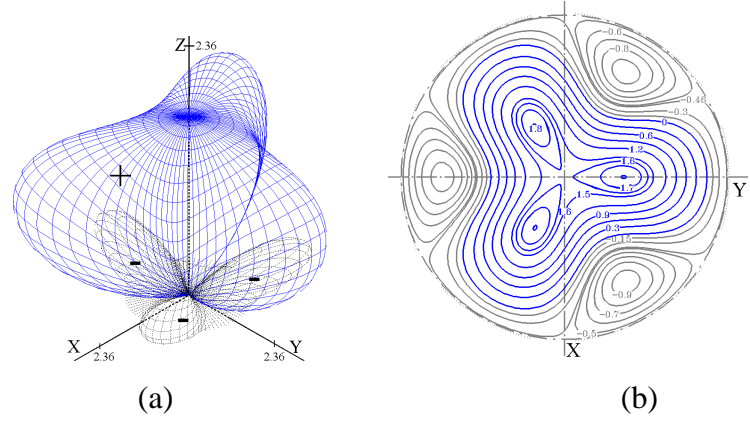

(a) (b)

Fig. 2. Half the indicative surface of the PO tensor (a) and its stereographic projection (b) for LiNbO$_3$ crystals (in the units of $pm^2/N$).

We have also constructed the indicative surfaces of that part of the PO tensor which is responsible for the effect occurring under the application of mechanical stresses $\sigma_{33}$ and $\sigma_{11}$. They are described respectively by the equations

(i) for $\sigma_{33}$:

$$R(\Theta,\varphi) = \pi_{13}\sin^2\Theta\cos^2\Theta + \pi_{33}\cos^4\Theta; \qquad (7)$$

(ii) for $\sigma_{11}$:

$$\begin{aligned}R(\Theta,\varphi) &= (\pi_{11}+\pi_{12})\sin^4\Theta\cos^2\varphi + \pi_{31}\sin^2\Theta\cos^2\Theta\cos^2\varphi \\ &\quad + \pi_{41}\sin 3\Theta\cos^2\varphi\sin\varphi\cos\Theta\end{aligned}. \qquad (8)$$

These surfaces are displayed in Fig. 3 and 4.

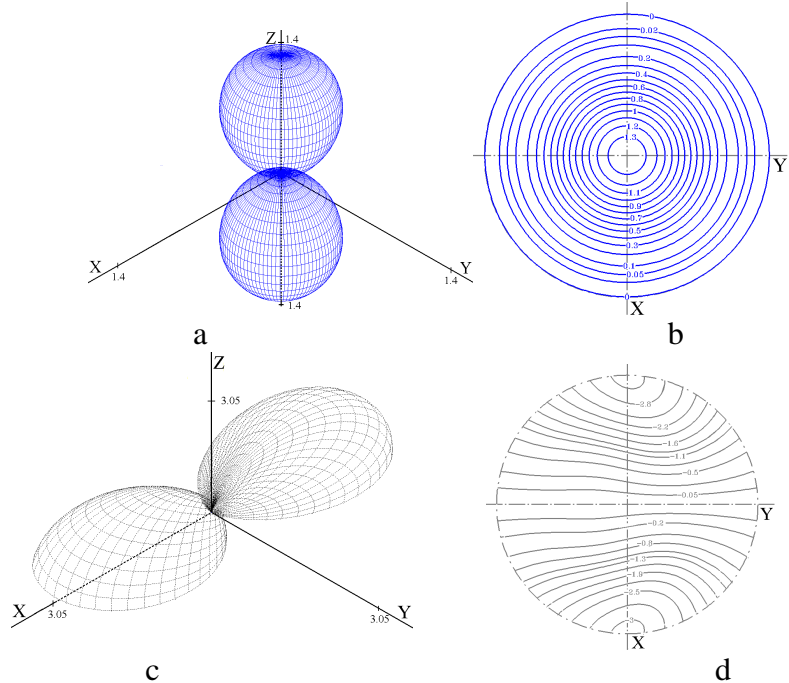

a b

c d

Fig. 3. Indicative surface of the part of PO tensor corresponding to the stress $\sigma_{33}$ (a), its stereographic projection (b), half the indicative surface of the part of PO tensor corresponding to the stress $\sigma_{11}$ (c) and its stereographic projection (d) for β-BaB$_2$O$_4$ crystals (in the units of $pm^2/N$).

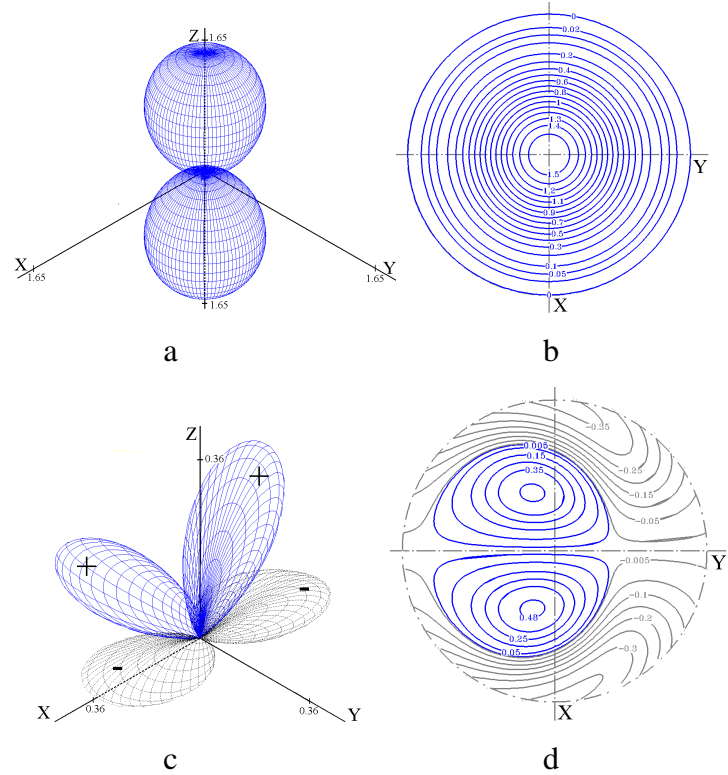

a  b

c  d

Fig. 4. Indicative surface of the part of PO tensor corresponding to the stress $\sigma_{33}$ (a), its stereographic projection (b), half the indicative surface of the part of PO tensor corresponding to the stress $\sigma_{11}$ (c) and its stereographic projection (d) for LiNbO$_3$ crystals (in the units of $pm^2/N$).

It is seen from Fig. 3 and 4 that the indicative surfaces belong to a lowered point symmetry group, which is actual under application of the chosen mechanical stress component. Since the point symmetry group of crystals remains to be *3m* when the component $\sigma_{33}$ is applied, the indicative surface also possesses all of the symmetry elements of the group *3m*. When the mechanical stress $\sigma_{11}$ is applied, the point symmetry group of crystal becomes *m* and so the symmetry of the indicative surfaces is also characterized by this group.

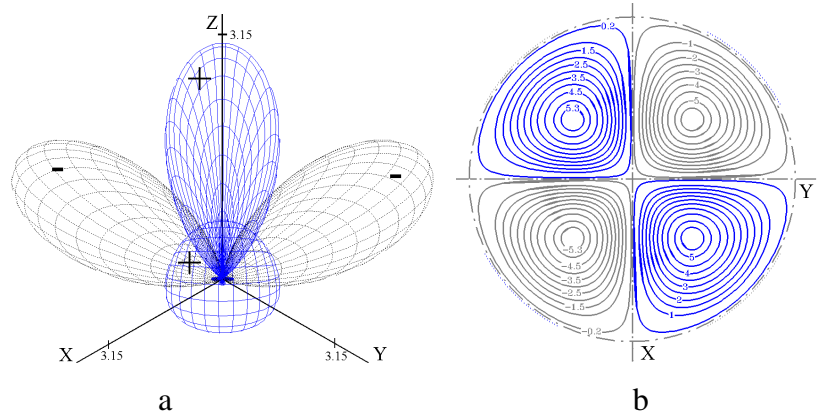

a  b

Fig. 5. Half the indicative surface of electrogyration tensor (a) and its stereographic projection (b) for BiGeO$_{20}$ crystals (in the units of $10^{-13}$ m/V).

Applicability of Eq. (2) for the construction of indicative surfaces of different higher-rank tensors could be demonstrated on the example of indicative surface of electrogyration tensor for BiGeO$_{20}$ crystals, which belong to the point symmetry group *23*. The components of the electrogyration tensor for these crystals are equal to $\gamma_{41} = \gamma_{52} = \gamma_{63} = 0.95\, pm/V$ [10]. The indicative surface of the electrogyration tensor for BiGeO$_{20}$ crystals is shown in Fig. 5.

**Conclusion**

We have reconstructed the surfaces of the PO tensor for β-BaB$_2$O$_4$ and LiNbO$_3$ crystals, which have been incorrectly presented in our previous papers. We thus demonstrate a correct approach for constructing the indicative surfaces of higher-rank tensors. The indicative surfaces of the PO effect published earlier (see [1-7]) are in fact the surfaces built in frame of the definition [8], though in a certain error. Moreover, characterization of piezoelectric effect by means of representation surfaces (but not indicative ones) in the work [8] has a well-defined practical meaning, in contrast to the case of PO effect.

**References**

1. Mytsyk B.H. and Andrushchak A.S. Kristallografiya **35** (1990) 1574 (in Russian).
2. Mytsyk B.H., Pryriz Ya.V. and Andrushchak A.S. Cryst. Res. Tech. **26** (1991) 931.
3. Mytsyk B.H. and Andrushchak A.S. Ukr. J. Phys. **38** (1993) 1015 (in Ukrainian).
4. Mytsyk B. Ukr. J. Phys. Opt. **4** (2003) 105.

5. Demyanyshyn N.M., Mytsyk B.H. and Kalynyak B.M. Ukr. J. Phys. **50** (2005) 519 (in Ukrainian).

6. Andrushchak A.S., Adamiv V., Krupych O., Martynyuk-Lototska I., Burak Y. and Vlokh R. Ferroelectrics **238** (2000) 299.

7. Vlokh O.G., Mytsyk B.H., Andrushchak A.S. and Pryriz Ya.V. Kristallografiya **45** (2000) 144 (in Russian).

8. Nye J.F. Physical properties of crystals. Clarendon Press, 1957.

9. Sirotin Yu.I. and Shaskolskaya M.P. Fundamentals of crystal physics, Moscow "Nauka" 1979 (in Russian).

10. Vlokh O.G. Spatial dispersion phenomena in parametric crystal optics, Lviv "Vyshcha shkola" 1984 (in Russian).